\documentclass[aps,pra,twocolumn,showpacs]{revtex4}

\newcommand{\pmin}{p_i^{\mbox{\small min}}}

\usepackage{graphicx}
\usepackage{txfonts}

\begin{document}

\title{Coherent control of a self-trapped Bose-Einstein condensate}

\author{C.E.~Creffield}
\affiliation{Department of Physics and Astronomy, University College London,
Gower Street, London WC1E 6BT, United Kingdom}

\date{\today}

\begin{abstract}
We study the behavior of a Bose-Einstein condensate held
in an optical lattice. We first show how a self-trapping transition
can be induced in the system by either increasing the number
of atoms occupying a lattice site, or by raising the interaction
strength above a critical value. We then investigate how applying
a periodic driving potential to the self-trapped state can be used
to coherently control the emission of a precise number of correlated
bosons from the trapping-site.  This allows the formation and transport 
of entangled bosonic states, which are of great relevance to novel 
technologies such as quantum information processing. 
\end{abstract}


\maketitle

{\em Introduction}
Ultracold atoms held in optical lattices are currently at the center of
intense theoretical and experimental investigation.
The experimental parameters of these systems can be controlled
extremely precisely, and in addition, the high degree of
isolation from the environment permits their quantum dynamics to remain
coherent over long timescales. Consequently such systems provide
an attractive way of investigating quantum many-body physics,
and it has been proposed \cite{toolbox} to use them to simulate strongly
correlated quantum systems of interest in other areas of physics
such as high-temperature superconductivity \cite{htsc},
They also provide an excellent starting
point for engineering and manipulating entangled states, which
are vital for implementations of quantum information processing. 

Bosons confined in an optical lattice provide an almost 
ideal realization \cite{jaksch} of the Bose-Hubbard (BH) model,
and a recent pioneering experiment directly observed
the quantum phase transition 
between a superfluid and a Mott insulator \cite{greiner}. 
Theoretical work \cite{eckardt,creffield} has shown how this
transition may be induced in an alternative way: by applying an
additional oscillatory driving field to suppress the inter-site
tunneling by means of the quantum interference effect
termed coherent destruction of tunneling (CDT) \cite{hanggi}.
In this work we show how such an oscillating driving field
can be used to manipulate the dynamics
of a different ground-state of the BH model -- the
{\em self-trapped} state. In particular we demonstrate 
an effect analogous to photon-assisted tunneling,
in which certain driving frequencies
induce a coherent oscillation of an integer number
of bosons between the trapping-site and its nearest-neighbors.
Combining this effect with CDT allows us to control the emission
of a definite number of bosons from the trapping site and manipulate
their speed of propagation through the optical lattice,
and thus enables the self-trapped state to be
used as a quantum beam-splitter or as a coherent source of entangled bosons. 

{\em Static properties}
The BH model is described by the Hamiltonian
\begin{equation}
H_{BH} = -J \sum_{\langle j, k \rangle} \left[ a_j^{\dagger} a_k + H.c. \right]
+ \frac{U}{2} \sum_j n_j \left( n_j - 1 \right),
\label{bh-ham}
\end{equation}	
where $a_j / a_j^{\dagger}$ are the standard annihilation/creation
operators for a boson on site $j$, 
$n_j = a_j^{\dagger} a_j$ is the number operator,
$J$ is the tunneling amplitude between neighboring sites, and $U$ is the
repulsion between a pair of bosons occupying the same site.  
Here we initialize the system in a state in which 
all the bosons occupy a {\em single} lattice site. As the Hubbard 
interaction is repulsive, it might be thought that such a state would 
be extremely unstable.
Surprisingly, however, this is not necessarily the case
and, depending on the strength of the interaction
and the filling of the lattice site, this highly-localized configuration 
is able to persist for long times.
In such cases the bosons are said to be ``self-trapped'' \cite{self_trap}.

Self-trapping has been recently observed experimentally in Bose-Einstein
condensates of roughly $1000$ atoms \cite{trap_expt,abietz},
and can be understood qualitatively by an energetics argument.
The presence of the optical lattice causes the energy
spectrum of non-interacting bosons to be confined to a Bloch band
of width $4J$. Consequently, if the potential energy per particle
of the trapped condensate is much higher than this, it cannot be
converted into the kinetic energy of free bosons and the
trapped state cannot decay. Its stability thus depends critically
on the absence of dissipative processes in optical lattice systems.

\begin{center}
\begin{figure}
\includegraphics[width=0.4\textwidth,clip=true]{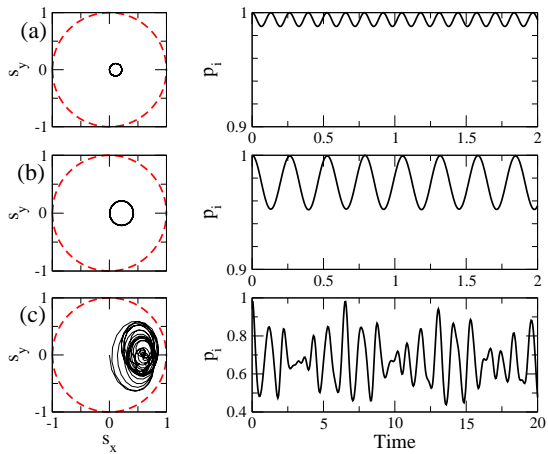}
\caption{Time evolution of a two-site system containing
7 bosons. The system is initialized in the state $|7,0\rangle$ 
which corresponds to the north pole of the Bloch sphere, and
evolves under the full BH Hamiltonian (\ref{bh-ham}).
To the left we show the evolution on the Bloch sphere, projected
onto the $(s_x, s_y)$ plane, and to the right the
overlap of the system with the initial state $p_i(t)$.
(a) Strong interaction, $U/J=8$. The system is described well
by the two-level model (\ref{2-level}), and
periodically traces out a circle on the
Bloch sphere. Correspondingly, $p_i$ makes high-frequency 
oscillations of small amplitude, 
and never attains a value smaller than 0.99, indicating
that self-trapping is occurring.
(b) Intermediate interaction $U/J=4$. The radius of the circular
orbit is larger, but the two-level approximation is still good.
The oscillations in $p_i$ are now of larger amplitude showing
that the trapping effect is reduced.
(c) Weak interaction, $U/J=1$. The two-level approximation now breaks down,
and the system's time-evolution is correspondingly more complicated. The
value of $p_i$ rapidly drops as the trapping effect has now
been completely lost.}
\label{bloch}
\end{figure}
\end{center}

To describe this effect quantitatively, we consider a 2-site BH model
holding $N$ bosons. If the system is initialized in the state
$|N,0 \rangle$ (that is, a Fock state with $N$ bosons occupying
one site with the other site empty), then the primary tunneling process
will be to the state $|N-1,1\rangle$. If we truncate the
Hilbert space to just these two states, we obtain an
effective two-level model
\begin{equation}
H_{\mbox{2-lev}} = \left( 
\begin{array}{cc}
V(N) & J \sqrt{N} \\
J \sqrt{N} & V(N-1)
\end{array}
\right),  
\label{2-lev}
\end{equation}
where $V(n)=U n (n-1)/2$ is the potential energy of $n$ bosons
occupying one site. It is useful to visualize the time evolution of
the system geometrically by making use of the Bloch sphere 
representation. Parameterizing $H_{\mbox{2-lev}}$ in terms of
the Pauli matrices
\begin{equation}
H_{\mbox{2-lev}} = \frac{U (N-1)^2}{2} \ I + J \sqrt{N} \  \sigma_x
+ \frac{U (N-1)}{2} \ \sigma_z ,
\label{2-level}
\end{equation}
reveals that we can interpret it as an
interaction between the Bloch vector ${\underline \sigma}$ and
a fictitious magnetic field ${\underline B}=(J \sqrt{N},0,U(N-1)/2)$.
Thus under the influence of the Hamiltonian the Bloch vector
will simply make a Larmor orbit about ${\underline B}$.
This form of time evolution is shown in Fig.\ref{bloch}a for
a strongly interacting ($U=8J$) 7-boson system. 
It can be clearly seen that, as expected, the Bloch vector traces out a
periodic circular orbit centered on ${\underline B}$.
Since for these parameters ${\underline B}$ is almost parallel
to the $s_z$ axis, the radius of this orbit is rather small,
and thus the system exhibits a high degree of self-trapping.

To assess the degree of self-trapping more precisely
we measure the overlap of the system's state with
the initial state, $p_i(t) = |\langle \psi_i | \psi(t) \rangle|^2$,
since when self-trapping occurs this quantity is restricted 
to values close to unity. The circular motion
shown in Fig.\ref{bloch}a equates to a very small-amplitude sinusoidal 
oscillation of $p_i$.
As $U$ is reduced, the angle between ${\underline B}$ and
the $s_z$ axis increases, and as a result the radius
of the Larmor orbit made by the Bloch vector increases (Fig.\ref{bloch}b).
Consequently the degree of trapping is reduced, and the amplitude
of oscillations of $p_i$ is larger.
If $U$ is decreased even further (Fig.\ref{bloch}c) the self-trapping
effect is lost, and the two-level approximation breaks down.
In this case the Bloch vector rapidly
leaves the surface of the Bloch sphere,
and its erratic time-evolution corresponds to an
irregular quasi-periodic behavior of $p_i$, which can
take very low values.

\begin{center}
\begin{figure}
\includegraphics[width=0.40\textwidth,clip=true]{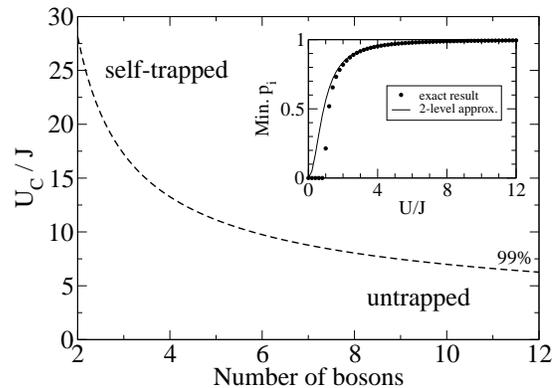}
\caption{The dashed-line plots the value of the Hubbard-interaction,
$U_C$, at which the dynamics of an undriven two-site system crosses over
from the untrapped to the self-trapped regime (Eq.\ref{boundary}).
For large particle numbers $U_C \sim 1/\sqrt{N}$, but rises
sharply as the number of bosons is reduced.
Inset: Comparison of the two-level approximation with the full BH Hamiltonian
for a 7-boson system. For $U/J < 1$ the 2-level approximation breaks down, 
but becomes increasingly accurate as $U$ is increased.}
\label{trap}
\end{figure}
\end{center}

In the inset of Fig.\ref{trap} we plot the minimum value of $p_i$
attained in the 7-boson system as a function of the interaction
strength. For small values of $U$ no self-trapping occurs,
and thus $\pmin$ takes a value of zero. As $U$ is increased,
however, the oscillations in $p_i$ are quenched, and for $U/J > 3$ 
it can be seen that the effective
two-level model describes the dynamics extremely well.
From Eq.\ref{2-level} we can obtain a criterion for the crossover
to the self-trapped regime, by defining the
transition to occur when $\pmin$ drops below a value $\alpha$. This yields
a value for the critical value of U,
\begin{equation}
\frac{U_C}{J} = \frac{2}{N  - 1} \sqrt{ \frac{N \alpha}{1-\alpha}}.
\label{boundary}
\end{equation}
This boundary is plotted in Fig.\ref{trap} for $\alpha=0.99$ -- that is,
when less than 1\% of the initial state is able
to leak to the other site. Although the self-trapping
regime is more easily achievable for large boson numbers since
$U_c \sim 1/\sqrt{N}$, it is in principle possible in a
system of just two bosons, if the interaction strength
can be raised sufficiently high. This has recently
been achieved experimentally \cite{pairing} in a gas
of trapped rubidium atoms.

{\em Dynamical properties} 
We now consider dynamically controlling the self-trapped state 
by applying a harmonic driving potential
\begin{equation}
H(t) = H_{BH} +  K \sin \omega t \sum_j \ j \  n_j ,
\label{drive}
\end{equation}
where $K$ is the amplitude of the driving field,
and $\omega$ is its frequency. 
We first consider the case of extracting a {\em single} boson
from the trapping site. As the occupation of the trap site changes
from $N$ to $N-1$ there is a corresponding loss of potential
energy $\Delta E = V_N - V_{N-1} = U(N-1)$, where $V_n$ is defined as
in Eq.\ref{2-lev}. When a system possesses such a large energy
gap, Floquet analysis may be used to show \cite{creffield,gp}
that extremely fine control over the tunneling dynamics is possible
at {\em multi-photon resonances}, that is, when $m \omega = \Delta E,
\ m=1,2 \dots$.
In general when this condition is satisfied, the system is
able to exchange energy with the driving field to overcome
the energy gap, and so tunneling is restored \cite{eckardt2}.
However, at particular values of the amplitude of the driving field,
CDT will occur when the Floquet quasienergies of the system
become degenerate, and correspondingly the dynamics of the system
will be frozen. For a sinusoidal driving potential these degeneracies
occur at the zeros of ${\cal J}_m(K/\omega)$, the $m$th Bessel
function of the first kind. 
Thus at a photon resonance it is possible to produce dramatic
differences in the tunneling rate by making small changes in
the amplitude of the driving field to move the system between 
CDT and photon assisted tunneling.

\begin{center}
\begin{figure}
\includegraphics[width=0.4\textwidth,clip=true]{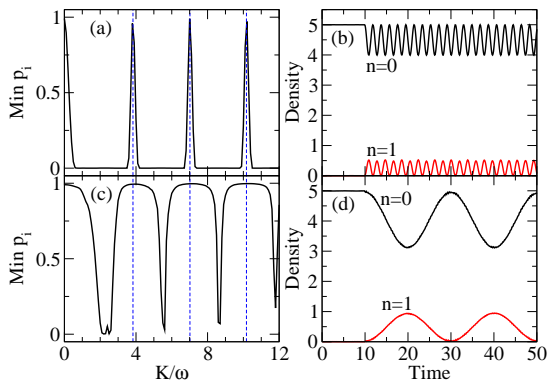}
\caption{Response of a 5-site system, holding 5 bosons
($U=16$), to a sinusoidal driving field.
(a) For a driving frequency $\omega = 4 U$, the value of
$\pmin$ rapidly drops as $K$ is increased from zero,
indicating that the initial state is rapidly destroyed
by photon-assisted tunneling.
At a sequence of sharp peaks centered on the
zeros of ${\cal J}_1(K/\omega)$ (vertical dashed lines)
coherent destruction of tunneling (CDT) instead causes the
system to be frozen in its initial state.
(b) Time dependence of the occupation numbers of the
lattice sites. For $t < 10$, $K/\omega$ is set
to be the first zero of ${\cal J}_1$, and CDT suppresses
any oscillations. For $t > 10$ we set $K/\omega=2.40$
which produces a sinusoidal oscillation of a single boson
from the central site $(n=0)$ to its two neighbors $(n=\pm 1)$.
(c) As in (a) but for a frequency of $\omega=3 U$. 
Again $\pmin$ peaks at the zeros of ${\cal J}_1(K/\omega)$,
although the peaks are rather broader.
(d) As in (b), but for $\omega=3 U$.
When $K/\omega$ switches to a value of 2.40
the number densities show a slow sinusoidal oscillation, in
which the occupation of the central site varies between five and three, i.e.
this driving frequency induces a {\em two-boson} oscillation.}
\label{4plot}
\end{figure}
\end{center}

In Fig.\ref{4plot}a we plot $\pmin$ obtained in a 
5-boson system driven at a frequency of $\omega=4 U$. 
This corresponds to the $m=1$ photon resonance. For $K=0$ the system
is self-trapped, and consequently $p_i$ remains near unity.
Increasing $K$, however, causes the value of $\pmin$
to rapidly drop to zero, demonstrating how photon assisted
tunneling overcomes the self-trapping effect.
As $K$ is increased further, $\pmin$ exhibits
a number of extremely sharp peaks centered on $K/\omega = 3.83,7.01$ and
$10.17$ -- the zeros of ${\cal J}_1(K/\omega)$. 

\begin{center}
\begin{figure}
\includegraphics[width=0.4\textwidth,clip=true]{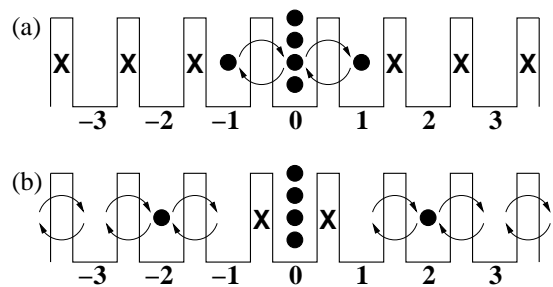}
\caption{(a) Schematic version of the process illustrated in
Fig.\ref{4plot}b. The driving field excites a boson from the
trapping site ($n=0$). If ${\cal J}_0(K/\omega) = 0$, CDT suppresses the
single boson tunneling processes (marked by X), and so
a Rabi-like oscillation occurs between the trapping site and its neighbors.
(b) Changing the driving parameters to stabilize the self-trapping
isolates the remaining bosons in the trapping site, but the
escaped particle is able to propagate through the optical lattice
with a renomalized tunneling amplitude $J_{\mbox{eff}}$.}
\label{scheme}
\end{figure}
\end{center}

Away from these zeros, the driving field causes
a single boson to diffuse symmetrically from the trapping site
to its neighboring lattice sites, and from there to
continue propagating through the optical lattice with a
renormalized tunneling \cite{hanggi}
$J_{\mbox{eff}} = J {\cal J}_0(K/\omega)$.
If we therefore choose a value of $K$ such that
$J_{\mbox{eff}} = 0$, the particle will not be able to propagate
further, and will thus just
make a Rabi-like oscillation between the trapping site
and its neighbors. This situation is illustrated schematically
in Fig.\ref{scheme}a.  We show in Fig.\ref{4plot}b the time dependence
of the occupation of the trapping site and its neighbor. Initially
we set $K/\omega = 3.83$ (a zero of ${\cal J}_1$), and 
no tunneling occurs: for this value
of $K/\omega$ CDT reinforces the self-trapping effect.
At $t=10$ we alter the amplitude of the driving to
$K/\omega = 2.40$ (a zero of ${\cal J}_0$) and a very clear
particle oscillation occurs, in which the occupation of
the trapping site cycles between 5 and 4, and that of the neighboring
sites moves between 0 and 0.5. In this sense the
optical lattice is acting as a particle beam-splitter, dividing
a single boson into a superposition of left and right-moving
components.

We have so far considered the case of extracting a single boson
from the trapping site. We can however apply a similar
method to induce the emission of an integer number 
of bosons, $N_{em}$, by noting that the 
energy difference per particle takes the remarkably simple
form 
\begin{eqnarray}
\Delta E / N_{em} &=& [V_N - (V_{N-N_{em}} + V_{N_{em}})]/N_{em} \nonumber \\
            {  }  &=& U(N-N_{em})
\end{eqnarray}
The self-trapped state thus contains a ladder of equally-spaced
energy levels, separated by $U$, and so by driving the system at the correct
frequency we can induce the emission of a given number of particles.
In Fig.\ref{4plot}c we show the response of the 5-boson system
to a driving field of frequency $\omega=3 U$,
thereby inducing the emission of {\em two} bosons.
As before we can observe peaks in $\pmin$ centered
on the zeros of ${\cal J}_1(K/\omega)$ at which CDT occurs,
while between them photon-assisted tunneling causes $\pmin$
to take low values. Fig.\ref{4plot}d shows the effect of switching 
the amplitude of the driving field to a value of $K/\omega=2.40$. 
We can again see that this has the effect of inducing a Rabi-like 
oscillation between the trapping site and its neighbors, but in
this case the oscillation indeed consists of two bosons, and
has a longer period.

\begin{center}
\begin{figure}
\includegraphics[width=0.405\textwidth,clip=true]{fig5a}
\includegraphics[width=0.45\textwidth,clip=true]{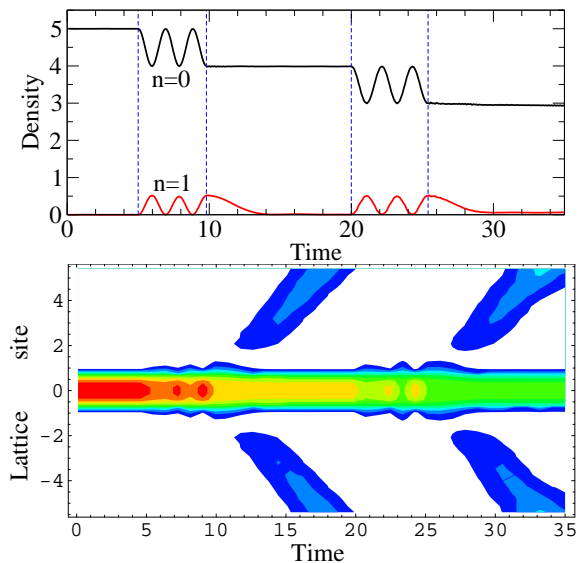}
\caption{Above: Time evolution of the particle density
of the trapping site ($n=0$) and its neighbor ($n=1$)
for a 5-boson system. The driving parameters are chosen
(see text) to produce the emission of single bosons
at $t\simeq 5$ and $25$, and thus the occupation of
the trapping site falls in steps as $5 \rightarrow 4 \rightarrow 3$.
Below: Particle density of the system. The emitted particle bursts
move away from the central site at a roughly constant speed,
and disperse slightly as they propagate through the lattice.} 
\label{emit}
\end{figure}
\end{center}

In Fig.\ref{emit} we show how combining the PAT effect with CDT allows
the population of the self-trapped state to be reduced step-by-step.
The system is initialized with 5 bosons in the central site,
and is initially driven at $\omega=4 U$. Driving the system
at $K/\omega=3.80$ induces CDT which suppresses the number variance
of the trapping site and stabilizes the self-trapped state.
At $t=5$ the value of $K/\omega$ is changed to 2.40 which induces
the Rabi-type oscillation between the trapping site and its
neighbors, as schematically shown in Fig.\ref{scheme}a.
After a {\em half-integer} number of these oscillations
we then alter the driving parameters
to $\omega=3 U, \ K/\omega=3.80$ which traps the remaining
4 bosons in the trapping site. For these parameters, however,
the single-particle tunneling is {\em not} quenched and so the
ejected boson is able to propagate through the optical lattice away 
from the trapping site, as shown in Fig.\ref{scheme}b. The emitted  
particle smears out to an extent
as it moves through the lattice (see Fig.\ref{emit}, lower panel),
but nonetheless the atom-pulse remains quite clearly defined
after propagating through several lattice spacings.
By repeating this procedure with appropriate driving frquencies
we can successively reduce the occupation of
the trapping site in integer steps, and thereby 
produce a sequence of well-defined, phase-coherent atom-pulses.

{\em Conclusions}
We have shown how self-trapping arises in a Bose-Einstein 
condensate confined to an optical lattice. 
Applying a resonant driving field to the self-trapped state
can either stabilize the trapping (when CDT occurs),
or can induce a Rabi-like particle oscillation.
The interplay between these effects
makes it possible to control the emission and propagation of
a precisely-defined number of particles, and thus
enables the self-trapped
state to be used as a particle beam-splitter
or a source of mesoscopic entangled states,
which have many possible applications in quantum information.

\bigskip
The author acknowledges numerous stimulating discussions
with Tania Monteiro. This work was supported by the EPSRC.


\begin{thebibliography}{99}
\bibitem{toolbox}
{D.~Jaksch and P.~Zoller, P. Ann. Phys. {\bf 315},
52 (2005).}

\bibitem{htsc}
{S.~Trebst, U.~Schollw\"ock, M.~Troyer, and P.~Zoller,
Phys. Rev. Lett. {\bf 96}, 250402 (2006).}

\bibitem{jaksch}
{D.~Jaksch, C.~Bruder, J.I.~Cirac, C.W.~Gardiner,
and P.~Zoller, Phys. Rev. Lett. {\bf 81}, 3108 (1998).}

\bibitem{greiner}
{M.~Greiner, O.~Mandel, T.~Esslinger, T.W.~H\"ansch,
and I.~Bloch, Nature {\bf 415}, 39 (2002).}

\bibitem{eckardt}
{A.~Eckardt, T.~Jinasundera, C.~Weiss, and M.~Holthaus,
Phys. Rev. Lett. {\bf 95}, 200401 (2005).}

\bibitem{creffield}
{C.E.~Creffield and T.S.~Monteiro, Phys. Rev. Lett. {\bf 96}, 210403
(2006).}

\bibitem{hanggi}
{F.~Grossmann, T.~Dittrich, P.~Jung, and P.~H\"anggi, Phys. Rev. Lett.
{\bf 67}, 516 (1991).}

\bibitem{self_trap}
{G.J.~Milburn, J.~Corney, E.M.~Wright, and D.F.~Walls,
Phys. Rev. A {\bf 55}, 4318 (1997).}

\bibitem{trap_expt}
{T.~Anker, M.~Abietz, R.~Gati, S.~Hunsmann, B.~Eiermann, A.~Trombettoni,
and M.K.~Oberthaler, Phys. Rev. Lett. {\bf 94}, 020403 (2005).}

\bibitem{abietz}
{M.~Albiez, R.~Gati, J.~F\"olling, S.~Hunsmann,
M.~Cristiani, and M.K.~Oberthaler, Phys. Rev. Lett. {\bf 95},
010402 (2005).}

\bibitem{pairing}
{K.~Winkler, G.~Thalhammer, F.~Lang, R.~Grimm, J.~Hecker~Denschlag,
A.J.~Daley, A.~Kantian, H.P.~Buchler, and P.~Zoller,
Nature {\bf 441}, 853 (2006).}

\bibitem{gp}
{C.E.~Creffield and G.~Platero, Phys. Rev. B {\bf 65}, 113304 (2002).}

\bibitem{eckardt2}
{A.~Eckardt, C.~Weiss, and M.~Holthaus,
Phys. Rev. Lett. {\bf 95}, 260404 (2005).}

\end{thebibliography}
\end{document}